\documentclass[conference]{IEEEtran}
\IEEEoverridecommandlockouts 
\usepackage{cite}
\usepackage{amsmath,amssymb,amsfonts}
\usepackage{algorithmic}
\usepackage{graphicx}
\usepackage{url}
\usepackage{textcomp}
\usepackage{xcolor}
\usepackage{changes}   % remove "final" to show changes  % za hlchange komandu
\usepackage{svg}

\def\BibTeX{{\rm B\kern-.05em{\sc i\kern-.025em b}\kern-.08em
    T\kern-.1667em\lower.7ex\hbox{E}\kern-.125emX}}
\begin{document}

\title{A Semantic Communication Approach to Fiducial Marker Processing in 5G-Enabled Edge SLAM\\

}

\author{\IEEEauthorblockN{Boris Radovanovic\IEEEauthorrefmark{1}, Vukan Ninkovic\IEEEauthorrefmark{1}\IEEEauthorrefmark{2}, Katarina Vidojevic\IEEEauthorrefmark{1}, Buda Bajic Papuga\IEEEauthorrefmark{1}, Dejan Vukobratovic\IEEEauthorrefmark{1} 
\vspace{1mm}
\IEEEauthorblockA{
\IEEEauthorblockA{\IEEEauthorrefmark{1}Faculty of Technical Sciences, University of Novi Sad, Serbia}
\IEEEauthorblockA{\IEEEauthorrefmark{2}The Institute for Artificial Intelligence Research and Development of Serbia, Serbia
}
}}}

\maketitle

\begin{abstract}
Autonomous robots increasingly rely on edge computing to offload 
computationally intensive perception tasks while maintaining real-time 
operation over 5G networks. However, conventional fiducial marker 
detection pipelines provide limited opportunities for efficient task 
partitioning, making them poorly suited for communication-aware edge 
deployment. This paper proposes a semantic split inference framework 
for fiducial marker processing in 5G-enabled Edge SLAM. A 
DeepTag-inspired convolutional neural network is partitioned between 
the robot and the edge server, where intermediate feature 
representations serve as task-oriented semantic information transmitted 
over the wireless link. The framework is integrated into a ROS2-based 
robotic architecture and characterized over a real 5G communication 
testbed. Experimental results demonstrate accurate keypoint estimation, 
illustrate the impact on downstream pose estimation, and quantify the 
communication--computation trade-offs associated with different split 
points, providing practical insights for communication-aware deployment 
of deep visual perception in connected robotic systems.
\end{abstract}


\section{Introduction}

The emergence of connected robotics is transforming the deployment of
autonomous systems across industrial automation, logistics, and smart
manufacturing environments~\cite{6Grob1}. By integrating robots with 5G and beyond
communication networks, computationally intensive perception and
decision-making tasks can be offloaded from resource-constrained
robotic platforms to nearby edge servers, thereby improving scalability, efficiency, and real-time operation while maintaining low
end-to-end latency~\cite{urllc, 6Gstd2}. In this context, reliable visual perception is essential for accurate localization and navigation of autonomous robots. Among the available approaches, fiducial marker-based simultaneous localization and mapping (SLAM)~\cite{slam} has emerged as a practical solution owing to its robustness, computational efficiency, and reliable pose estimation in structured indoor environments.

Despite the benefits of edge-enabled robotic perception, conventional
fiducial marker processing remains communication intensive. Existing implementations typically transmit raw camera images or fixed intermediate representations such as detected corners, resulting in communication
overhead that increases latency and limits scalability in
bandwidth-constrained deployments~\cite{bezerra}. Recent advances in deep learning have
demonstrated that learned keypoint regression networks can significantly
improve localization accuracy while naturally supporting split
inference, where the processing pipeline is partitioned between the
robot and the edge server~\cite{evgenidis_2025, xue_2025}. However, existing split inference approaches
primarily focus on computational partitioning, whereas the transmitted
intermediate representations inherently constitute task-oriented
semantic information that can be jointly optimized for efficient
communication and downstream perception. This perspective has received
limited attention in the context of fiducial marker processing for Edge
SLAM~\cite{BenAli2020, Xu2020}.

Motivated by these observations, this paper proposes a semantic split
inference framework for fiducial marker processing in 5G-enabled Edge
SLAM. A DeepTag-based keypoint regression network~\cite{deeptag} is employed as the
perception module, whose encoder can be partitioned at multiple
locations to generate intermediate representations with different
communication and computational characteristics. The proposed framework
is implemented within a ROS2-based \cite{ros2} robotic architecture and evaluated
on a real 5G-enabled edge computing testbed. Experimental results
demonstrate that accurate downstream localization can be maintained
while substantially reducing the communication overhead over the 5G
link, and analyze the communication--computation trade-offs associated
with different split points.

\section{Background}

\subsection{Fiducial Marker Processing Pipeline in SLAM}

%Fiducial markers have served for decades as a de facto standard in photogrammetry, camera calibration, and machine vision. Since their introduction, they have been widely adopted in robotics for tasks such as localization and calibration. A fiducial marker system typically comprises a predefined set of marker patterns together with an algorithm for their detection and validation; a wide variety of marker designs and detection algorithms have been proposed, each tailored toward efficient and/or robust detection. Regardless of the approach, in the context of the SLAM problem these detections are used to estimate the robot's pose, as illustrated in Fig. \ref{Fig_1}.

\begin{figure*}[!t]
\centering
\includegraphics[width=5in]{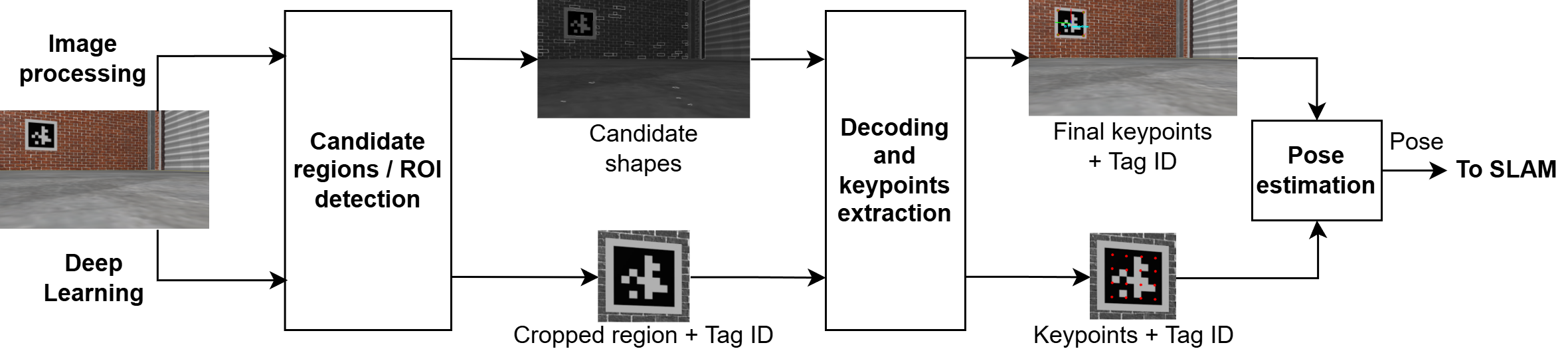}
\caption{Fiducial Marker Processing Pipeline in SLAM.}
\label{Fig_1}
\end{figure*}

Fiducial markers are widely used in robotics, photogrammetry, camera calibration, and machine vision for reliable pose estimation. A fiducial marker system consists of a predefined marker dictionary and a corresponding detection algorithm, with numerous designs proposed to improve detection efficiency and robustness. In fiducial marker-based SLAM, the detected markers provide geometric observations that enable robot localization and mapping, making the fiducial marker processing pipeline a fundamental component of the overall system, as illustrated in Fig.~\ref{Fig_1}.

\subsubsection{Fiducial Marker Processing via Image Processing}

For years, fiducial marker detection has relied on classical image-processing techniques \cite{artag}. These methods can be coarsely divided into a two-stage pipeline: first, candidate marker regions are detected and isolated; second, the marker payload within each candidate is decoded to determine its identity. The output of the pipeline is the set of image-space keypoints of each detected marker together with its decoded identifier. This is illustrated in Fig. \ref{Fig_1}, using the AprilTag 2 detector~\cite{apriltag2} as a representative example.

As shown in the figure, the candidate-detection stage operates on the full input image. For AprilTag 2, it comprises three steps: binarization of the input image via adaptive thresholding; connected-component segmentation and clustering of the resulting black-and-white regions; and the fitting of a quadrilateral to each resulting cluster of boundary pixels. The second stage then operates only at the few candidate locations: each candidate is decoded through an $\mathcal{O}(1)$ hash-table lookup, and optionally, its corners are refined to improve localization accuracy. The output of this stage is the set of valid tag detections, each represented by its four corner keypoints and its tag identifier.

A key advantage of classical fiducial marker detectors is their high computational efficiency, enabling real-time execution on resource-constrained devices. They also achieve very low false-positive rates through marker coding with a large minimum Hamming distance and decoding within a small Hamming radius (typically one or two bit errors). However, these pipelines provide no suitable point for edge offloading, since the data exchanged between successive stages remains comparable to the input image size and may even increase during intermediate processing (e.g., segmentation). Although the candidate-detection stage produces a compact output, offloading the remaining processing offers little benefit, as it consists primarily of identifier decoding and corner refinement. Furthermore, classical detectors exhibit limited robustness to marker occlusion and image degradation, such as motion blur.

\subsubsection{Fiducial Marker Processing via Deep Learning}

With the rapid advancement of deep learning (DL), convolutional neural networks (CNNs) have become the dominant approach for computer vision tasks such as localization, tracking, and fiducial marker detection~\cite{deepcharuco,deepformabletag}. Unlike classical pipelines, learned approaches are not constrained by predefined coding and decoding schemes, enabling custom marker designs and end-to-end optimization. DeepTag~\cite{deeptag} is a representative CNN-based framework for accurate fiducial marker detection that also supports custom markers. Similar to the classical pipeline, it consists of two stages: region-of-interest (ROI) detection and keypoint regression (Fig.~\ref{Fig_1}, lower part).

In the first stage, an ROI detection network processes the input image, identifies candidate marker regions, and decodes their identities. Each ROI is subsequently passed to a second network that estimates the marker keypoints. The final output consists of the image-space keypoints together with the corresponding marker identifier. Compared with conventional detection, learning-based approaches offer improved robustness to occlusion and image degradation while supporting split inference. The intermediate feature representations generated within the network provide multiple candidate split points, enabling different trade-offs between onboard computation, communication overhead, and edge-side processing.

%TODO: In the first stage, an ROI-detection network takes the full input image and predicts coarse regions likely to contain markers, producing a small set of candidate boxes. In the second stage, each candidate region is cropped and passed to a second network that localizes fine-grained keypoints (the marker's outer corners) and, in DeepTag's case, an internal grid of keypoints, and decodes the marker's identity. As in the classical pipeline, the final output is the set of image-space keypoints and the decoded identifier for each marker. Relative to the classical approach, learning-based detection offers markedly higher robustness to occlusion and image degradation and removes any dependence on a fixed marker family, at the cost of potentially higher computational demand. Importantly for edge-assisted SLAM, the two-stage learned architecture exhibits a property the classical pipeline lacks: the first stage can run on the robot and emit a compact intermediate representation, either the cropped ROIs or a learned bottleneck feature, that is small relative to the input image, while the heavier second stage is offloaded to the edge server. Whereas the classical pipeline achieves substantial payload reduction only after nearly all of its computation is complete, the learned pipeline places a natural bottleneck ahead of its most expensive stage, providing a convenient split point for offloading.

\subsection{5G-Enabled Edge SLAM}

 SLAM is a fundamental capability of autonomous mobile robots, enabling simultaneous estimation of the robot pose and reconstruction of an unknown environment from sensor observations. Modern visual SLAM algorithms rely on continuous streams of high-resolution sensor data and increasingly incorporate computationally demanding perception modules, such as feature extraction, semantic segmentation, object detection, and graph optimization. Executing these workloads entirely onboard resource-constrained robots can significantly limit operating time, sensing quality, and achievable autonomy.

\begin{figure*}[!t]
\centering
\includegraphics[width=4.69in]{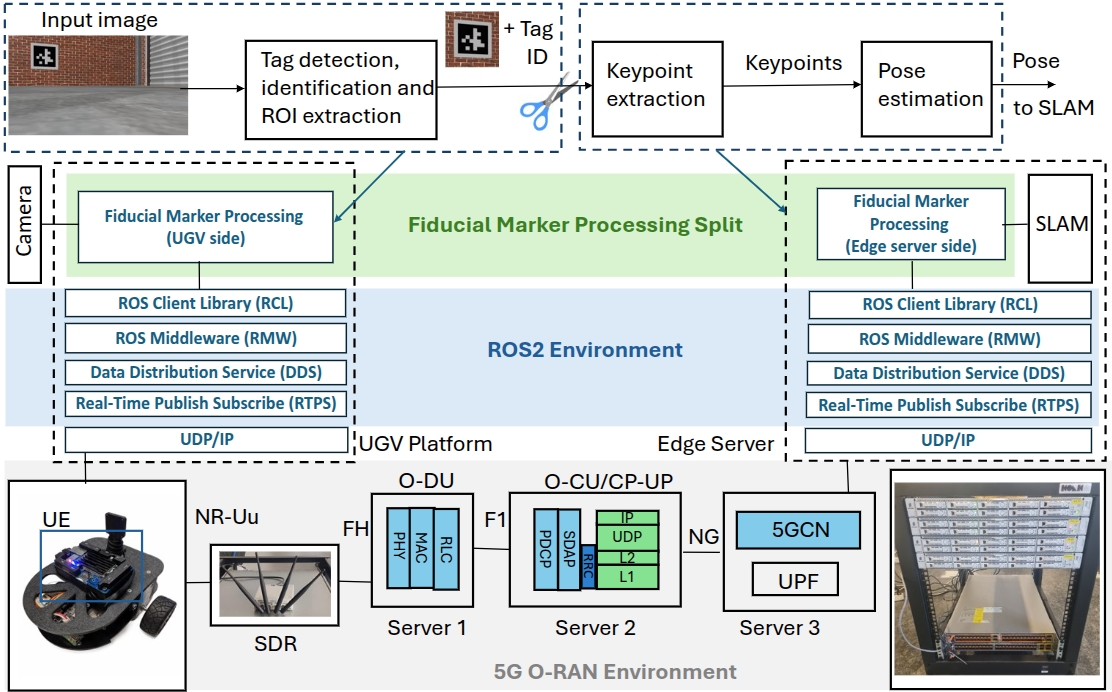}
\caption{Fiducial Marker Processing Pipeline in 5G-Enabled Edge SLAM.}
\label{Fig_2}
\end{figure*}

%To overcome these limitations, recent research has proposed Edge SLAM, where computationally intensive components of the SLAM pipeline are offloaded to edge servers over high-performance wireless networks. Ben Ali et al. introduced Edge-SLAM \cite{BenAli2020}, which dynamically partitions visual SLAM between the robot and the edge according to network and computational conditions, while Xu et al. proposed edge-assisted semantic visual SLAM \cite{Xu2020} by offloading semantic perception tasks to edge infrastructure. These studies demonstrate that edge computing can substantially improve the computational efficiency of mobile robots while preserving localization accuracy. However, they primarily focus on computation partitioning and task scheduling, assuming the communication network as an ideal transport mechanism and providing little insight into the practical impact of wireless communication on end-to-end SLAM performance.

To overcome these limitations, recent research has proposed Edge SLAM, where computationally intensive SLAM components are offloaded to edge servers over high-performance wireless networks. Ben Ali \emph{et al.}~\cite{BenAli2020} dynamically partition visual SLAM between the robot and the edge according to network and computational conditions, while Xu \emph{et al.}~\cite{Xu2020} offload semantic perception tasks to edge infrastructure. These studies demonstrate the potential of edge computing to improve robotic perception while preserving localization accuracy. However, they primarily focus on computation partitioning, treating the communication network as an ideal transport mechanism and providing limited insight into the impact of wireless communication on end-to-end SLAM performance.

%The emergence of private 5G networks and mobile edge computing (MEC) has made practical deployment of Edge SLAM increasingly feasible. Recent experimental studies have demonstrated end-to-end SLAM operation over both private industrial 5G deployments~\cite{Sosalla2025} and commercial 5G macro-cellular networks~\cite{Karfakis2023}, confirming that 5G connectivity can support offloading of visual perception tasks while maintaining real-time operation. Nevertheless, existing implementations largely treat the robotic middleware and the communication system as independent components. In particular, the interaction between ROS2 publish–subscribe communication, DDS/RTPS transport mechanisms, and the underlying 5G user plane remains insufficiently understood, despite its direct influence on latency, bandwidth utilization, and reliability.

The emergence of private 5G networks and mobile edge computing (MEC) has made practical deployment of Edge SLAM increasingly feasible. Recent studies have demonstrated end-to-end SLAM over both private industrial 5G deployments~\cite{Sosalla2025} and commercial 5G networks~\cite{Karfakis2023}, confirming that 5G can support offloading of visual perception tasks while maintaining real-time operation. Nevertheless, existing implementations largely treat the robotic middleware and communication system as independent components. As a result, the interaction between ROS2 publish--subscribe communication, DDS/RTPS transport, and the underlying 5G user plane remains largely unexplored, despite its direct impact on latency, bandwidth utilization, and reliability. Moreover, recent industrial developments have demonstrated the practical integration of ROS2 DDS communication with 5G infrastructures~\cite{EP2022}, further motivating communication-aware optimization of distributed robotic perception systems.

\section{Fiducial Marker Processing Pipeline in 5G-Enabled Edge SLAM}

%In this section, we present a detailed integration model for fiducial marker processing pipeline in the context of 5G-enabled edge SLAM implementation.

\subsection{System Model and Architecture}

Fig.~\ref{Fig_2} illustrates the proposed system architecture for 5G-enabled Edge SLAM based on fiducial marker processing. The system consists of three tightly integrated components: i) a mobile unmanned ground vehicle (UGV) equipped with a camera and a 5G user equipment (UE), ii) a 5G communication infrastructure compliant with the 3GPP/O-RAN architecture, and iii) an edge server executing the remaining stages of the visual processing pipeline and SLAM.

The software architecture follows the ROS2 distributed execution model, in which functional modules are implemented as ROS2 nodes communicating via the publish--subscribe paradigm. On the UGV, the camera node acquires image frames and forwards them to the fiducial marker processing node, which executes either the complete pipeline or only its initial stages before transmitting the intermediate representation to the edge server. The edge executes the remaining processing stages, estimates the relative six-degree-of-freedom (6D) camera-to-marker pose, and forwards it to the SLAM node for localization and mapping. Communication between the UGV and the edge is provided by the ROS2 middleware, where messages are serialized using DDS/RTPS and transported over UDP/IP through the 5G network.

Let the image acquired by the onboard camera be denoted by
$\mathbf{x}\in\mathbb{R}^{H\times W\times C}$, where $H$, $W$, and $C$
represent the image height, width, and number of channels, respectively.
The first stage of the fiducial marker processing pipeline detects and
identifies fiducial markers, producing a  ROI for
each detected marker (Fig.~\ref{Fig_1}) and yielding
$
\mathcal{R}=\{(\mathbf{r}_m,d_m)\}_{m=1}^{M},
$
where $M$ is the number of detected markers, $\mathbf{r}_m$ denotes the
ROI of the $m$-th marker, and $d_m$ its decoded identifier. The
corresponding ROIs are then processed by the keypoint extraction stage,
producing
$
\mathcal{K}=\{(\mathbf{k}_m,d_m)\}_{m=1}^{M},
$
where $\mathbf{k}_m$ denotes the set of image-space keypoints associated
with the $m$-th marker, including both the marker corners and, depending
on the detection algorithm, additional learned feature points. Finally,
the pose estimation stage computes
$
\mathcal{Y}=\{(\mathbf{T}_m,d_m)\}_{m=1}^{M},
$
where $\mathbf{T}_m$ denotes the relative 6D camera-to-marker pose.
Accordingly, the complete fiducial marker processing pipeline is
represented as
$
\mathcal{Y}=\mathcal{F}(\mathbf{x}).
$

The communication infrastructure follows the 3GPP/O-RAN architecture. The UE embedded within the UGV communicates over the NR-Uu interface with a disaggregated gNB comprising an O-DU and O-CU interconnected through the F1 interface. User-plane traffic is forwarded through the 5G Core Network (5GCN), including the User Plane Function (UPF), towards the edge server hosting the edge server side fiducial marker processing and SLAM application. From the viewpoint of the ROS2 application, the 5G system provides transparent end-to-end transport of ROS2 messages while simultaneously determining \emph{the communication latency that directly contributes to the overall task execution time}.

Unlike conventional Edge SLAM, where the primary optimization concerns the allocation of entire SLAM modules between the robot and the edge, this work focuses on the fiducial marker processing pipeline itself. As illustrated in Fig.~\ref{Fig_2}, the pipeline may be partitioned at different locations, resulting in different amounts of data transmitted over the 5G link. Let the fiducial marker processing pipeline consist of an ordered sequence of processing blocks
$\mathcal{P}=\{P_1,P_2,\ldots,P_N\}$, where each block $P_i$ implements the processing function
$f_i(\cdot)$. The pipeline progressively transforms the captured image
$\mathbf{x}$ into the marker pose set $\mathcal{Y}$ consumed by the SLAM
algorithm, and can therefore be expressed as
$\mathcal{F}=f_N\circ f_{N-1}\circ\cdots\circ f_1$.
A split point $k$, where $1\leq k<N$, partitions the pipeline into the
onboard subset
$\mathcal{P}_{UGV}=\{P_1,\ldots,P_k\}$ and the edge subset
$\mathcal{P}_{ES}=\{P_{k+1},\ldots,P_N\}$.

%Let the processing pipeline be represented by an ordered sequence of $processing blocks $\mathcal{P}=\{P_1, P_2, \ldots, P_N\}$, where the first blocks operate directly on the captured image and the final block produces the 6D pose estimate consumed by the SLAM algorithm. Equivalently, the pipeline can be expressed as the composition of processing block functions 
%$\mathcal{F}=f_N\circ f_{N-1}\circ\cdots\circ f_1.$ A split point $k$ partitions the pipeline into two subsets: $\mathcal{P}_{UGV}=\{P_1, P_2, \ldots, P_k\}$ and $\mathcal{P}_{ES}=\{P_{k+1}, P_{k+2}, \ldots, P_N\}$, executed on the UGV and the edge server, respectively. The intermediate representation generated after block $P_{k}$ is transmitted through the ROS2/5G communication stack.

Accordingly, the processing functions executed on the UGV and edge server are defined as
$\mathcal{F}_{UGV}=f_k\circ\cdots\circ f_1$ and
$\mathcal{F}_{ES}=f_N\circ\cdots\circ f_{k+1}$,
respectively, such that
$\mathcal{F}=\mathcal{F}_{ES}\circ\mathcal{F}_{UGV}$.
The onboard subsystem generates the intermediate representation
$\mathbf{z}=\mathcal{F}_{UGV}(\mathbf{x})$,
which is transmitted to the edge server through the ROS2/5G communication stack, where the final output is computed as
$\mathcal{Y}=\mathcal{F}_{ES}(\mathbf{z})$.
\emph{The optimal split depends on the computational capabilities of the UGV and edge server, the wireless network characteristics, and the size of the transmitted representation.} Unlike conventional image-processing pipelines, which produce compact outputs only after most computations are completed, DL-based pipelines naturally support split inference by transmitting intermediate feature representations that preserve task-relevant information.

%\begin{figure*}[!t]
%\centering
%\includegraphics[width=5.5in]{Figure_3_CSCN_2026.jpg}
%\caption{\textcolor{red}{Testbed and Implementation setup.}}
%\label{Fig_3}
%\end{figure*}

\subsection{A Semantic Communications Approach}

The partitioned processing model  (Section~III-A) naturally establishes the connection between split inference and semantic communications~\cite{semantic_split}. The onboard function generates the intermediate representation $\mathbf{z}=\mathcal{F}_{UGV}(\mathbf{x})$, which serves as the input to the edge-side function $\mathcal{F}_{ES}$. Since the network is trained end-to-end, $\mathcal{F}_{UGV}$ learns a latent representation that retains only the information required for downstream pose estimation rather than reconstructing the input image. Consequently, $\mathbf{z}$ constitutes a task-oriented semantic representation that suppresses redundant information while preserving features necessary for accurate marker pose estimation.

This property fundamentally changes the role of the communication subsystem. In conventional Edge SLAM, transmitted information typically consists of raw images or handcrafted visual descriptors whose representation is independent of the downstream task. In contrast, the proposed framework communicates the learned latent representation $\mathbf{z}$, explicitly optimized for the edge-side processing function $\mathcal{F}_{ES}$. Consequently, the communication objective shifts from reliable bit-level transmission to preserving the task-relevant information required for accurate downstream inference.

From a system perspective, the overall performance is jointly determined by the selected split point and the semantic representation transmitted over the wireless link. The split point defines the onboard processing subset $\mathcal{P}_{UGV}=\{P_1,P_2,\ldots,P_k\}$, thereby determining both the distribution of computational workload between the UGV and the edge server and the characteristics of the intermediate representation $\mathbf{z}$. Consequently, it directly affects the onboard computational complexity, communication overhead, and, ultimately, the delay and accuracy of the downstream localization task.

\section{Implementation and Experimental Results}

\subsection{DeepTag Implementation and Training}

\subsubsection{Network Architecture}

The proposed DeepTag implementation follows an encoder--regressor
architecture that supports the split inference framework introduced in
Section~III. As discussed in Section~II-A, the first stage relies on the
classical AprilTag~2 detector, which detects fiducial markers, decodes
their identifiers, and extracts an ROI around each detected tag. This
stage is not learned in the present work, although fully learned
two-stage detectors~\cite{deeptag} and end-to-end architectures remain
promising directions for future work.

The network takes as input a grayscale ROI of size $256\times256$
pixels and predicts the normalized coordinates of $K=16$ image-space
keypoints arranged as a $4\times4$ grid. It consists of a VGG-style
convolutional encoder followed by a regression head. The encoder
comprises five convolutional blocks, each containing two $3\times3$
convolutional layers followed by batch normalization, ReLU activation,
and $2\times2$ max-pooling, progressively reducing the feature-map
dimensions ($H\times W\times C$) as
$
256\times256\times1
\rightarrow
128\times128\times32
\rightarrow
64\times64\times64
\rightarrow
32\times32\times128
\rightarrow
16\times16\times256
\rightarrow
8\times8\times256.
$
The regression head consists of a $3\times3$ convolutional layer,
adaptive average pooling, and two fully connected layers
($4096\!\rightarrow\!512\!\rightarrow\!32$). The complete network
contains approximately $5.06\times10^6$ trainable parameters. Each encoder block defines a candidate split point, whose output feature map serves as the intermediate representation $\mathbf{z}$ transmitted to the edge server. The resulting communication--computation trade-offs are evaluated in Section~IV-B.

\subsubsection{Dataset Generation}
Since sub-pixel ground-truth keypoint annotations are difficult to obtain from real imagery, the network is trained primarily on a synthetic dataset with exact labels available by construction. Each sample is generated by rendering a planar AprilTag from the \texttt{tag36h11} family under a randomly sampled six-degree-of-freedom (6-DoF) pose using a calibrated camera model matching the onboard camera ($f_x\!\approx\!f_y\!\approx\!1045$\,px, $1280\times720$ resolution). Camera orientation and distance are sampled independently to cover diverse viewing conditions, while the corresponding ground-truth keypoints are obtained by exact perspective projection.

To improve robustness, the rendered images are augmented using random brightness and contrast variations, Gaussian blur, additive Gaussian noise, gamma correction, and randomized background textures. In addition, the detected tag quadrilateral is randomly perturbed before warping to the fixed $256\times256$ network input, following the Stage-2 training strategy of DeepTag~\cite{deeptag}. The resulting dataset comprises 10,000 samples, randomly partitioned into training and test subsets using a 70/30 split.

To reduce the domain gap, the network is subsequently fine-tuned on 8,000 photorealistic images rendered in Gazebo Harmonic. The simulated environment contains complete 3D scenes with realistic lighting, materials, and viewpoints while retaining pixel-accurate annotations. AprilTag markers are placed at known poses and observed from randomly sampled camera viewpoints, with ground-truth keypoints obtained by exact projection. The dataset is randomly partitioned into training and test subsets using a 70/30 split.

\subsubsection{Training}
Training is performed in two stages. The network is first trained from scratch on the synthetic dataset and subsequently fine-tuned on the rendered dataset to improve generalization to realistic deployment conditions. Training employs a spatially weighted variant of the Wing loss~\cite{wingloss}, which places greater emphasis on small localization errors than the mean-squared-error (MSE) loss and is therefore well suited to keypoint regression. The base loss is defined as:
\begin{equation}
\ell(\delta)=
\begin{cases}
w\ln\!\left(1+\frac{|\delta|}{\epsilon}\right), & |\delta|<w,\\
|\delta|-C, & \text{otherwise},
\end{cases}
\label{eq:wing}
\end{equation}
where $\delta$ denotes the coordinate-wise prediction error, $w=10$, $\epsilon=2$, and $C=w-w\ln(1+w/\epsilon)$ ensures continuity.

To reduce the influence of boundary keypoints on downstream PnP pose estimation, each keypoint is assigned a weight according to its normalized distance from the nearest patch edge,
$d_{\mathrm{edge}}=\min_{c\in\{x,y\}}\min(k_c,1-k_c)$,
where $k_c$ denotes the ground-truth normalized coordinate. The weight is computed as
$\omega=\mathrm{clamp}(d_{\mathrm{edge}}/0.5,0.3,1.0)$,
yielding the training objective:
\begin{equation}
\mathcal{L}
=
\frac{1}{KB}
\sum_{b=1}^{B}
\sum_{k=1}^{K}
\omega_k^{(b)}
\,
\ell
\left(
\hat{\mathbf{k}}_k^{(b)}
-
\mathbf{k}_k^{(b)}
\right),
\end{equation}
where $B$ is the mini-batch size, $K=16$ is the number of keypoints, and $\hat{\mathbf{k}}_k^{(b)}$ and $\mathbf{k}_k^{(b)}$ denote the predicted and ground-truth normalized coordinates of the $k$-th keypoint in the $b$-th training sample, respectively. The network is optimized using Adam~\cite{Kingma:2014aa} with cosine annealing for 70 epochs, a batch size of $32$, and initial learning rates of $3\times10^{-4}$ and $5\times10^{-5}$ during pretraining and fine-tuning, respectively.

\subsection{Communication Performance}

\begin{figure}[!t]
\centering
\includegraphics[width=2.2in]{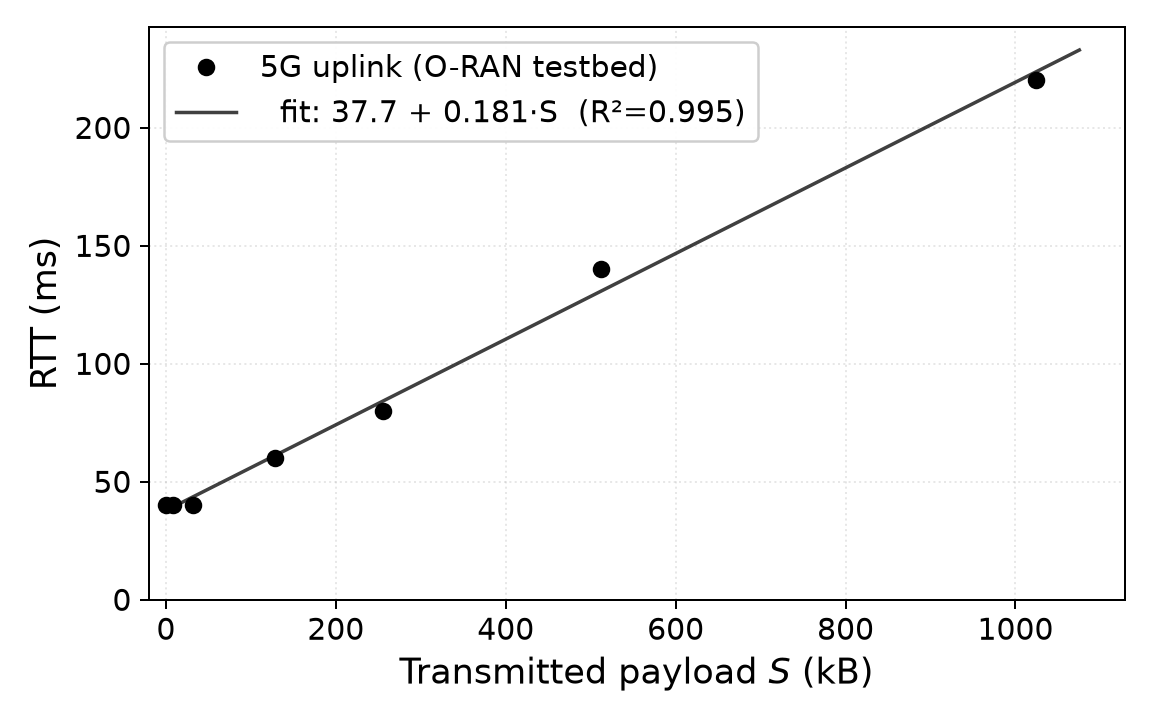}
\caption{Uplink RTT vs. payload size, with fitted link model.}
\label{Fig_link_model}
\end{figure}

The measurements were performed on the 5G O-RAN testbed described in \cite{balkancom2026}, based on a 20~MHz bandwidth srsRAN/Open5GS setup. The communication link was characterized by transmitting synthetic payloads sized to each candidate split point and measuring round-trip time (RTT) over 200 repetitions per size. As RTT is measured on a single clock, it requires no time synchronization between the UGV and the server, and the results can be described by the model:

\begin{equation}
\mathrm{RTT} = t_0 + \beta \cdot S,
\label{eq:rtt_model}
\end{equation}
where $t_0$ is payload-independent floor, $\beta$ the marginal cost per kilobyte, and $S$ the payload size (Fig. \ref{Fig_link_model}). 

A least-squares fit over the full payload range gives $\beta = 0.18$~ms/kB. We report the measured floor latency, $t_0 = 39.9$~ms, rather than the fitted intercept ($37.7$~ms), since the link is not perfectly linear. Payloads below ${\sim}32$~kB exhibit the same latency, indicating a radio scheduling floor rather than transfer time. Consequently, further payload reduction provides no additional latency benefit. Increasing the allocated uplink resources (by balancing uplink/downlink TDD slots) improved $\beta$ by a factor of ${\sim}3$ ($14.2 \rightarrow 43.5$~Mbps) while leaving $t_0$ unchanged.

%A least-squares fit over the full payload range gives $\beta = 0.18$~ms/kB. We report $t_0 = 39.9$~ms as measured directly at sub-kilobyte payloads rather than as the fitted intercept ($37.7$~ms), since the link is not perfectly linear. The measured floor value is attributable to radio scheduling granularity rather than transfer time, as payloads below ${\sim}32$ KB incur identical latency regardless of size. Consequently, further payload reduction yields no additional latency benefit. \textcolor{red}{Increasing the allocated uplink resources (by balancing uplink/downlink TDD slots) improved $\beta$ by a factor of ${\sim}3$ ($14.2 \rightarrow 43.5$~Mbps) while leaving $t_0$ unchanged.}

\subsection{Perception and Localization Performance}

\subsubsection{Perception Performance}
\begin{figure}
    \centering
    \includegraphics[width=0.37\linewidth]{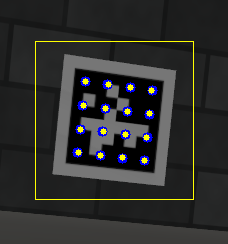}
    \caption{Representative keypoint regression result -  Ground-truth (blue) and predicted (yellow) keypoints.}
    \label{fig:test_sample}
\end{figure}

The perception performance of the proposed keypoint regression network is evaluated on a held-out test set unseen during training. Prediction accuracy is quantified by the mean absolute pixel error,
$e_{\mathrm{px}}=\frac{S}{K}\sum_{i=1}^{K}\|\hat{\mathbf{k}}_i-\mathbf{k}_i\|_2$,
where $S=256$ is the patch size and $K=16$, and the relative keypoint error,
$e_{\mathrm{rel}}=(e_{\mathrm{px}}/L_{\mathrm{tag}})\times100\%$,
where $L_{\mathrm{tag}}=S\|\mathbf{k}_{\max}-\mathbf{k}_{\min}\|_2$ is the projected diagonal of the ground-truth keypoint grid. The proposed network achieves a mean absolute keypoint error of $1.47$\,px ($\sigma=0.79$\,px), corresponding to a mean relative keypoint error of $0.90\%$ ($\sigma=0.47\%$), demonstrating accurate and consistent keypoint estimation under diverse viewing conditions. Fig.~\ref{fig:test_sample} shows a representative test sample, illustrating the close agreement between the predicted and ground-truth keypoint locations across the entire $4\times4$ grid.

\subsubsection{Localization Performance}
To assess the impact of the predicted keypoints on downstream localization, the representative test sample from Fig.~\ref{fig:test_sample} is processed using a PnP algorithm to estimate the relative camera-to-marker pose. The estimated pose is compared with the simulator ground truth using the translation error,
$e_t=\|\hat{\mathbf{t}}-\mathbf{t}\|_2$,
and the rotation error,
$e_r=\arccos\!\left((\mathrm{tr}(\hat{\mathbf{R}}\mathbf{R}^{\top})-1)/2\right)$,
where $(\hat{\mathbf{t}},\hat{\mathbf{R}})$ and $(\mathbf{t},\mathbf{R})$ denote the estimated and ground-truth translation and rotation, respectively. The proposed approach achieves a translation error of $0.83$\,cm and a rotation error of $2.04^\circ$. Using the same PnP solver, the conventional AprilTag pipeline yields a translation error of $1.52$\,cm and a rotation error of $0.37^\circ$. Since the primary objective of this work is to validate the proposed semantic split inference framework, localization performance is illustrated on a representative example, while the perception module is evaluated over the complete test set.

\subsubsection{Split Point Analysis}

Table \ref{tab:blocks_pld_cmpt} lists, for each candidate split point, the size of the transmitted representation and the cumulative on-board compute required to produce it, profiled single-threaded on an x86 CPU (Intel Core i7-1165G7) without GPU acceleration. Compute accumulates progressively while payload decreases monotonically, so that the two costs trade-off against one another.

\begin{table}[htbp]
\caption{Payload size and cumulative compute at each split point.}
\footnotesize
\setlength{\tabcolsep}{4pt}
\renewcommand{\arraystretch}{1.05}
\begin{centering}
\begin{tabular}{|c|c|c|}
\hline
\textbf{Split point} $k$ & \textbf{Payload $S(k)$ [kB]} & \textbf{Cumulative compute $C(k)$ [ms]} \\
\hline
Block 1 & 1024 & 13.6 \\
Block 2 & 512  & 24.9  \\
Block 3 & 256 & 33.7 \\
Block 4 & 128 & 43.3 \\
Block 5 & 32  & 47.3  \\
AvgPool & 8 & 48.1 \\
Full (keypoints) & 0.06  & 48.3  \\
\hline
\end{tabular}
\label{tab:blocks_pld_cmpt}
\end{centering}
\end{table}

Combining the measured link model (Eq.~\eqref{eq:rtt_model}) with the per-block compute profile yields the end-to-end cost of each candidate split:

\begin{equation}
T(k) = \gamma \cdot C(k) + t_0 + \beta \cdot S(k),
\label{eq:split-cost}
\end{equation}
where $\gamma$ is a scaling factor, introduced to evaluate the optimum across a range of on-board compute capabilities, since $C(k)$ is at present profiled only on an x86 CPU rather than on the UGV's embedded processor.

\begin{figure}[!t]
\centering
\includegraphics[width=2.5in]{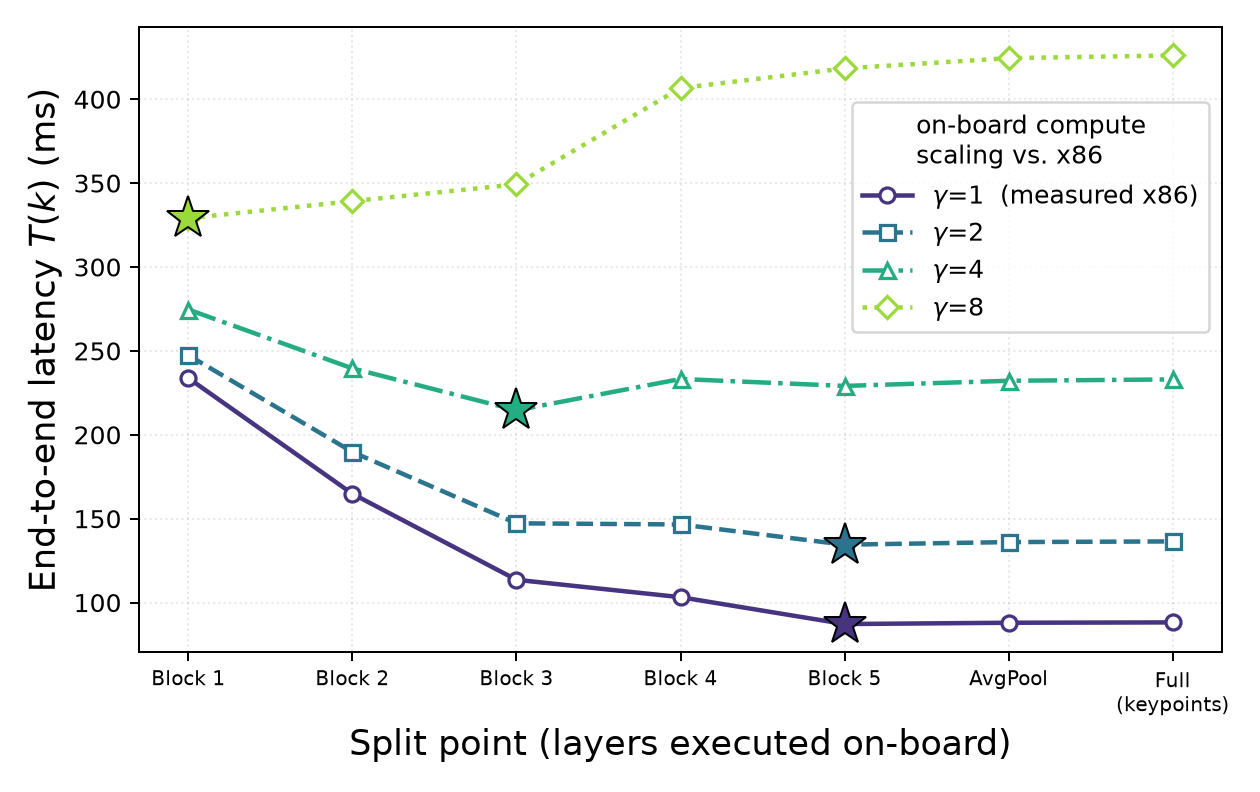}
\caption{Split point cost on the 5G uplink ($\star$ = optimum).}
\label{Fig_split_cost}
\end{figure}

Fig.~\ref{Fig_split_cost} evaluates Eq.~\eqref{eq:split-cost} using the measured 5G uplink. The end-to-end cost exhibits a clear minimum whose location depends on the onboard computational capability. For $\gamma\leq2$, the optimum lies at the deepest layers, as transmitting a $256$~kB representation requires ${\sim}80$~ms while offloading the remaining layers saves only ${\sim}14$~ms of computation. For $\gamma\geq4$, the optimum shifts toward earlier layers. These results motivate adaptive semantic partitioning instead of a fixed split.

\section{Conclusions and Future Work}
This paper presented a semantic split inference framework for fiducial 
marker processing in 5G-enabled Edge SLAM, in which a DeepTag-inspired 
keypoint regression network is partitioned between the robot and the 
edge server. Intermediate feature representations are interpreted as 
task-oriented information, enabling communication-aware 
deployment over ROS2 and 5G. Experimental results confirmed accurate 
keypoint estimation and quantified communication--computation 
trade-offs at different split points. Future work will investigate 
adaptive split selection under dynamic network conditions and 
joint optimization of the semantic representation and wireless 
transmission for end-to-end robotic perception.

\section*{Acknowledgments}
Supported by the Serbian Ministry of Science, Technological Development and Innovation (Project No. 00101957 2025 13440 003 000 620 021) and China’s National Key R\&D Program (Grant 2024YFE0197400).


\begin{thebibliography}{00}

\bibitem{6Grob1}
M. Ghassemian \emph{et al.},
``6G Empowering Future Robotics: A Vision for Next-Generation Autonomous Systems,''
\emph{IEEE Commun. Mag.}, 2026.

%\bibitem{6Gstd1}
%M. Ghassemian \emph{et al.},
%``Standardisation Landscape for 6G Robotic Services,''
%in \emph{Proc. IEEE Conf. Standards Commun. Netw. (CSCN)},
%2023, pp. 148--154.

\bibitem{urllc}
D. E. Boubiche \emph{et al.},
``The Next Generation of Internet of Robotic Things (IoRT): Leveraging 6G for Intelligence, Connectivity, and Scalability,''
\emph{IEEE Internet Things Mag.}, early access.


\bibitem{6Gstd2}
IEEE Std. P1955,
``Standard for 6G Empowering Robotics.''
[Online]. Available: \url{https://standards.ieee.org/ieee/1955/11660/}
%\bibitem{Kabiri2024}
%M. Kabiri \emph{et al.},
%``Graph-Based vs. Error State Kalman Filter-Based Fusion of 5G and Inertial Data for MAV Indoor Pose Estimation,''
%\emph{J. Intell. Robot. Syst.},
%vol. 110, no. 2, Art. no. 87, 2024.

\bibitem{slam}
H. Durrant-Whyte and T. Bailey,
``Simultaneous Localization and Mapping: Part I,''
\emph{IEEE Robot. Autom. Mag.},
vol. 13, no. 2, pp. 99--110, 2006.

\bibitem{bezerra}
R. Bezerra \emph{et al.},
``AI-IoT-Robotics Integration: Survey of Frameworks, Emerging Trends, and the Path Toward Connected Robotics,''
\emph{IEEE Internet Things J.},
vol. 13, no. 10, pp. 20398--20412, 2026.

\bibitem{evgenidis_2025}
N. G. Evgenidis \emph{et al.},
``Split Learning in Computer Vision for Semantic Segmentation Delay Minimization,''
\emph{IEEE J. Sel. Areas Commun.},
vol. 43, no. 12, pp. 3955--3968, 2025.

\bibitem{xue_2025}
C. Xue \emph{et al.},
``Adaptive Multi-Robot Cooperative Localization Based on Distributed Consensus Learning of Unknown Process Noise Uncertainty,''
\emph{IEEE Trans. Autom. Sci. Eng.},
vol. 22, pp. 8738--8761, 2025.

%\bibitem{cao_2024}
%H. Cao, S. Shreedharan, and N. Atanasov,
%``Multi-Robot Object SLAM Using Distributed Variational Inference,''
%\emph{IEEE Robot. Autom. Lett.},
%vol. 9, no. 10, pp. 8722--8729, Oct. 2024.

\bibitem{BenAli2020}
A. J. Ben Ali, Z. S. Hashemifar, and K. Dantu,
``Edge-SLAM: Edge-Assisted Visual Simultaneous Localization and Mapping,''
in \emph{Proc. 18th Int. Conf. Mobile Syst., Appl., Services (MobiSys)},
2020, pp. 325--337.

\bibitem{Xu2020}
J. Xu \emph{et al.}, ``Edge assisted mobile semantic visual SLAM,'' in \emph{Proc. IEEE Conf. Comput. Commun. (INFOCOM)}, 2020, pp. 1828--1837.

\bibitem{deeptag}
Z. Zhang, Y. Hu, G. Yu, and J. Dai, ``DeepTag: A General Framework for Fiducial Marker Design and Detection,'' \emph{IEEE Trans. Pattern Anal. Mach. Intell.}, vol. 45, no. 3, pp. 2931--2944, 2023.

\bibitem{ros2}
S. Macenski \emph{et al.},
``Robot Operating System 2: Design, Architecture, and Uses in the Wild,''
\emph{Science Robotics}, vol. 7, no. 66, 2022.

%\bibitem{artoolkit}
%\textcolor{red}{H. Kato and M. Billinghurst, ``Marker Tracking and HMD Calibration for a Video-Based Augmented Reality Conferencing System,'' in \emph{Proc. IEEE/ACM Int. Workshop Augmented Reality (IWAR)}, 1999, pp. 85--94.}

\bibitem{artag}
M. Fiala, ``ARTag, a Fiducial Marker System Using Digital Techniques,'' in \emph{Proc. IEEE Comput. Soc. Conf. Comput. Vis. Pattern Recognit. (CVPR)}, vol. 2, pp. 590--596, 2005.

%\bibitem{aruco}
%S. Garrido-Jurado, R. Mu\~{n}oz-Salinas, F. J. Madrid-Cuevas, and M. J. Mar\'{i}n-Jim\'{e}nez,
%``Automatic Generation and Detection of Highly Reliable Fiducial Markers Under Occlusion,''
%\emph{Pattern Recognit.}, vol. 47, no. 6, pp. 2280--2292, 2014.

\bibitem{apriltag2}
J. Wang and E. Olson, ``AprilTag 2: Efficient and Robust Fiducial Detection,'' in \emph{Proc. IEEE/RSJ Int. Conf. Intell. Robots Syst. (IROS)}, 2016, pp. 4193--4198

\bibitem{deepcharuco}
D. Hu, D. DeTone, and T. Malisiewicz, ``Deep ChArUco: Dark ChArUco Marker Pose Estimation,'' in \emph{Proc. IEEE/CVF Conf. Comput. Vis. Pattern Recognit. (CVPR)}, 2019, pp. 8436--8444.

%\bibitem{e2etag}
%J. B. Peace, E. Psota, Y. Liu, and L. C. Pérez, ``E2ETag: An End-to-End Trainable Method for Generating and Detecting Fiducial Markers,'' in \emph{Proc. Brit. Mach. Vis. Conf. (BMVC)}, 2020.

\bibitem{deepformabletag}
M. B. Yaldiz, A. Meuleman, H. Jang, H. Ha, and M. H. Kim, ``DeepFormableTag: End-to-End Generation and Recognition of Deformable Fiducial Markers,'' \emph{ACM Trans. Graph.}, vol. 40, no. 4, pp. 1--14, 2021.

%\bibitem{6Grob2}
%``6G Empowering Future Robotics: Mapping Requirements and Advancements for the IMT-2030 Framework,'' One6G White Paper, available at: https://one6g.org/resources/publications/, Mar. 2025. 


%\bibitem{6Gmrs1}
%Z. Chen, K.C. Chen, C. Dong, and Z. Nie, ``6G mobile communications for multi-robot smart factory,'' \emph{J. ICT Standardization,} vol. 9, no. 3, pp. 371-404, 2021.

%\bibitem{6Gmrs2}
%J. Bravo-Arrabal, R. Vázquez-Martín, J.J. Fernández-Lozano, and A. García-Cerezo, ``Strengthening multi-robot systems for SAR: co-designing robotics and communication towards 6G,'' arXiv preprint arXiv:2504.01940, 2025.

%\bibitem{6Gts1}
%M. Dohler, S. Saikali, A. Gamal, M.C. Moschovas, and Y. Patel, ``The crucial role of 5G, 6G, and fiber in robotic telesurgery,'' \emph{J. Robot. Surg.}, vol. 19, no. 1, p.4, 2024.

%\bibitem{6Gts2}
%N.A. Mohammedali, T. Kanakis, and M.O. Agyeman, ``Survey on the Impact of AI, Robotics and 6G Networks on the Remote Surgery,'' in \emph{Proc. Int. Conf. Softw., Telecommun. Comput. Netw. (SoftCOM)}, pp. 1-6. Sept. 2024.

%\bibitem{5Grt1}
%I. Harjula, M. Uitto, M. Jurmu, J. Koskinen, J. Mäkelä, S. Walter, M. Hentula, T. Heikkilä, M. Lintala, and K. Rautiola, ``Smart manufacturing multi-site testbed with 5g and beyond connectivity,'' in \emph{Proc. IEEE Int. Symp. Pers., Indoor Mobile Radio Commun. (PIMRC),}  Sept. 2021.

%\bibitem{PRbook}
%S. Thrun, W. Burgard, D. Fox, ``Probabilistic robotics,'' MIT Press, 2005.


%\bibitem{Ge2025}
%Z. Ge \emph{et al.}, ``Sensing with Mobile Devices through Radio SLAM: Models, Methods, Opportunities, and Challenges,'' \emph{IEEE Communications Magazine,} 63(12), pp.80-87, 2025.

%\bibitem{Macenski2022}
%S. Macenski, T. Foote, B. Gerkey, C. Lalancette, W. Woodall, ``Robot operating system 2: Design, architecture, and uses in the wild,'' \emph{Science robotics,} 7(66), 2022.

%\bibitem{orantut}
%M. Polese, L. Bonati, S. D’Oro, S. Basagni, and T. Melodia, ``Understanding O-RAN: Architecture, interfaces, algorithms, security, and research challenges,'' \emph{IEEE Commun. Surveys Tuts.,} vol. 25, no. 2, pp. 1376-1411, 2025.

%\bibitem{Bouknana2025}
%N. Bouknana, M. Ahadi, F. Kaltenberger, R. Schmidt, ``An O-RAN Framework for AI/ML-Based Localization with OpenAirInterface and FlexRIC,'' arXiv preprint arXiv:2511.19233, 2025.

%\bibitem{Polese2026}
%M. Polese, R. Gangula, T. Melodia, ``Enabling Programmable Inference and ISAC at the 6GR Edge with dApps,'' arXiv preprint arXiv:2603.29146, 2026.

%\bibitem{srsran_project} Software Radio Systems, srsRAN Project - Open Source RAN, Available at: https://www.srsran.com/5g [Accessed: 20/12/25].

%\bibitem{open5gs} Open5GS - Open source NR/LTE mobile network, Available at: https://open5gs.org/ [Accessed: 20/12/25].


\bibitem{Sosalla2025}
P. Sosalla, ``Multi-Access Edge Computing for Mobile Robots,'' Ph.D. dissertation, Technische Universität Dresden, Dresden, Germany, 2025.

\bibitem{Karfakis2023}
P. T. Karfakis, M. S. Couceiro, and D. Portugal, ``NR5G-SAM: A SLAM framework for field robot applications based on 5G New Radio,'' \emph{Sensors}, vol. 23, no. 11, Art. no. 5354, 2023.

\bibitem{EP2022}
eProsima, ``eProsima and Ericsson simplify 5G integration in ROS 2,'' 2022. [Online]. Available:
https://www.eprosima.com/news/eprosima-and-ericsson-simplify-5g-integration-in-ros2.
Accessed: Apr. 2026.

\bibitem{semantic_split}
J. Choi \emph{et al.}, "Semantics Alignment via Split Learning for Resilient Multi-User Semantic Communication," \emph{IEEE Trans. Veh. Technol.,} vol. 73, no. 10, pp. 15815-15819, 2024.

\bibitem{wingloss}
Z.-H. Feng \emph{et al.},
``Wing Loss for Robust Facial Landmark Localisation With Convolutional Neural Networks,''
in \emph{Proc. IEEE Conf. Comput. Vis. Pattern Recognit. (CVPR)}, 2018, pp. 2235--2245.

\bibitem{Kingma:2014aa}
D.~P. Kingma and J.~L. Ba, ``Adam: A method for stochastic optimization,'' in
\emph{Proc. Int. Conf. Learn. Representation (ICLR),} 2015,
pp.~1--41.

\bibitem{balkancom2026}
B.~Radovanovic, S.~Talosi, S.~Sobot, and D.~Vukobratovic,
``CSI-Assisted Edge SLAM Testbed Platform for 5G Connected Unmanned Autonomous Vehicles,''
\emph{arXiv preprint arXiv:2607.10394}, 2026.


\end{thebibliography}
\end{document}